\documentclass[usenatbib]{mn2e}

\usepackage{psfig}
\usepackage{graphicx}
\usepackage{subfig}
\usepackage{ textcomp }
\usepackage{amsmath}
\usepackage{amssymb}
\usepackage{natbib}
\usepackage{gensymb}
\usepackage{color}
\usepackage{caption}

\newcommand{\RNum}[1]{\uppercase\expandafter{\romannumeral #1\relax}}
\def\gs{\mathrel{\raise0.35ex\hbox{$\scriptstyle >$}\kern-0.6em\lower0.40ex\hbox{{$\scriptstyle \sim$}}}}
\def\ls{\mathrel{\raise0.35ex\hbox{$\scriptstyle <$}\kern-0.6em\lower0.40ex\hbox{{$\scriptstyle \sim$}}}}

\def\Wm2{\,\hbox{W}\,\hbox{m}^{-2}}
\def\gsim{\mathrel{\raise0.35ex\hbox{$\scriptstyle >$}\kern-0.6em\lower0.40ex\hbox{{$\scriptstyle \sim$}}}}
\def\lsim{\mathrel{\raise0.35ex\hbox{$\scriptstyle <$}\kern-0.6em\lower0.40ex\hbox{{$\scriptstyle \sim$}}}}

\begin{document}

\title[Molecular gas in FIR-luminous QSOs at {\it z}\,$\sim$\,2.5]{
The evolutionary connection between QSOs and SMGs: molecular gas in far-infrared luminous QSOs at {\it z}\,$\sim$\,2.5}

\author[Simpson et al.]
{ \parbox[h]{\textwidth}{ 
J.\,M.~Simpson,$^{\! 1,*}$ 
Ian~Smail,$^{\! 1}$
A.\,M.~Swinbank,$^{\! 1}$ 
D.\,M.~Alexander,$^{\! 1}$ 
R.~Auld,$^{\! 2}$
M.~Baes,$^{\! 3}$
D.\,G.~Bonfield,$^{\! 4}$
D.\,L.~Clements,$^{\! 5}$
A.~Cooray,$^{\! 6}$ 
K.\,E.\,K.~Coppin,$^{\! 7}$ 
A.\,L.\,R~Danielson,$^{\! 1}$ 
A.~Dariush,$^{\! 2,5}$
L.~Dunne,$^{\! 8}$
G.~de Zotti,$^{\! 9,10}$ 
C.\,M.~Harrison,$^{\! 1}$
R.~Hopwood,$^{\! 5}$
C.~Hoyos,$^{\! 11}$
E.~Ibar,$^{\! 12}$
R.\,J.~Ivison,$^{\! 12,13}$ 
M.\,J.~Jarvis,$^{\! 4,14}$ 
A.~Lapi,$^{\! 10,15}$ 
S.\,J.~Maddox,$^{\! 8}$
M.\,J.~Page,$^{\! 16}$ 
D.\,A.~Riechers,$^{\! 17}$ 
E.~Valiante,$^{\! 2}$
P.\,P.~van\,der\,Werf\,$^{\! 18}$
}
\vspace*{8pt}\\
$^1$Institute for Computational Cosmology, Department of Physics, Durham University, South Road, Durham DH1 3LE, UK\\
$^2$School of Physics and Astronomy, Cardiff University, Queen's Buildings, The Parade, Cardiff CF24 3AA, UK  \\
$^3$Sterrenkundig Observatorium, Universiteit Gent, Krijgslaan 281 S9, B-9000 Gent, Belgium\\
$^4$Centre for Astrophysics Research, Science \& Technology Research Institute, University of Hertfordshire, Hatfield, AL10 9AB, UK\\
$^5$Astrophysics Group, Blackett Lab, Imperial College, Prince Consort Road, London SW7 2AZ, UK\\
$^6$Department of Physics and Astronomy, University of California, Irvine, CA 92697, USA \\
$^7$Department of Physics, McGill University, 3600 Rue University, Montr\'eal, QC, H3A 2T8, Canada \\
$^8$Dept of Physics and Astronomy, University of Canterbury, Private Bag 4800, Christchurch 8140, New Zealand \\
$^9$INAF-Osservatorio Astronomico di Padova, Vicolo Osservatorio 5, I-35122 Padova, Italy\\
$^{10}$SISSA, Via Bonomea 265, I-34136 Trieste, Italy \\
$^{11}$School of Physics and Astronomy, University of Nottingham, University Park, Nottingham NG7 2RD, UK\\
$^{12}$UK Astronomy Technology Centre, Royal Observatory, Blackford Hill, Edinburgh EH9 3HJ\\
$^{13}$Institute for Astronomy, University of Edinburgh, Blackford Hill, Edinburgh EH9 3HJ\\
$^{14}$Physics Department, University of the Western Cape, Cape Town, 7535, South Africa\\
$^{15}$Dipartimento di Fisica, Universit\`{a} “Tor Vergata,” Via della Ricerca Scientifica 1, 00133 Roma, Italy\\
$^{16}$Mullard Space Science Laboratory, University College London, Holmbury St Mary, Dorking, Surrey RH5 6NT, UK\\
$^{17}$California Institute of Technology, MC 249-17, 1200 East California Boulevard, Pasadena, CA 91125, USA \\
$^{18}$Leiden Observatory, Leiden University, P.O. Box 9513, 2300 RA Leiden, The Netherlands\\
$^*$email: j.m.simpson@dur.ac.uk\\
}

\maketitle

\begin{abstract} 
We present IRAM Plateau de Bure Interferometer observations of the $^{12}$CO\,(3--2) emission from two far-infrared luminous QSOs at $z\sim $\,2.5 selected from the {\it Herschel}-ATLAS survey.  These far-infrared bright QSOs were selected to have supermassive black holes (SMBHs) with masses similar to those thought to reside in sub-millimetre galaxies (SMGs) at $z\sim$\,2.5; making them ideal candidates as systems in transition from an ultraluminous infrared galaxy phase to a sub-mm faint, unobscured, QSO.  We detect $^{12}$CO\,(3--2) emission from both QSOs and we compare their baryonic, dynamical and SMBH masses  to those of SMGs at the same epoch.   We find that these far-infrared bright QSOs have similar dynamical but lower gas masses than SMGs.  In particular we find that far-infrared bright QSOs have $\sim$\,50\,$\pm$\,23\% less warm/dense gas than SMGs, which combined with previous results showing the QSOs lack the extended, cool reservoir of gas seen in SMGs, suggests that they are at a different evolutionary stage. This is consistent with the hypothesis that far-infrared bright QSOs represent a short ($\sim 1$\,Myr) but ubiquitous phase in the transformation of dust obscured, gas-rich, starburst-dominated SMGs into unobscured, gas-poor, QSOs. 
\end{abstract}


\section{Introduction}\label{sec:intro}

An evolutionary link between local ultraluminous infrared galaxies (ULIRGs, with far-infrared [FIR] luminosities of L$_{\rm FIR}\geq 10^{12}$\,L$_\odot$) and QSOs\footnote{QSO, quasi-stellar object, defined as having M$_{\rm B } \leq -22$ and one or more broad emission lines with a width $\geq$ 1000kms$^{-1}$} was first suggested by \citet{Sanders88} and considerable effort has been expended to test this connection in the local Universe (e.g.\ \citealt{Tacconi02,Veilleux09}).  This hypothesis has been strengthened by the discovery of a relation between the mass of supermassive black holes (SMBHs) and the mass of their host spheroids (e.g.\ \citealt{Magorrian98,Gebhardt00,Ferrarese00}), which suggests a physical connection between the growth of spheroids and their SMBHs.  As shown by theoretical simulations, such a link can be formed through the suppression of star formation in a host galaxy by winds and outflows from an active galactic nucleus (AGN) at its centre \citep{DiMatteo05,Hopkins05}.  

Both the ULIRG and QSO populations evolve very rapidly with redshift, and although subject to selection biases, intriguingly both populations appear to reach a broad peak in their activity at $z\sim 2.5$ \citep{Shaver88, Chapman05, Wardlow11}. The high-redshift ULIRGs, which are typically bright at sub-millimetre wavelengths, and hence are called sub-millimetre galaxies (SMGs), are thought to represent the formation of massive stellar systems through an intense burst of star formation in the early Universe, triggered by mergers of gas-rich progenitors~\citep{Frayer98,Blain02,Swinbank06b,Swinbank10,Tacconi06,Tacconi08, Engel10}. 

The similarity in the redshift distribution of the SMGs and QSOs may be indicating that the evolutionary link between these populations, which has been postulated locally, also holds at $z\sim 2$, when both populations were much more numerous, and significant in terms of stellar mass assembly. Indeed recent work on the  clustering of $z\sim 2$ QSOs and SMGs has shown that  both populations reside in parent halos of a similar mass,  M$_{\rm halo} \sim 10^{13}$\,M$_\odot$ \citep{Hickox12}, adding another circumstantial connection between them.  Moreover, this characteristic M$_{\rm halo}$ is  similar to the mass at which galaxy populations transition from star forming to passive systems (e.g.\ \citealt{Brown08,Coil08,Hartley10}) which may be indirect evidence for the influence of QSO-feedback on star formation in galaxies~\citep{Granato01,Lapi06}.  One other piece of circumstantial evidence for a link between SMGs and QSOs comes from estimates of the masses of SMBHs in SMGs. Due to the dusty nature of SMGs these studies are inherently challenging, yielding results with more scatter than seen for optical quasars. However, they suggest that the SMBH masses are $\sim10^8$\,M$_\odot$, significantly lower than seen in comparably massive galaxies at the present-day \citep{Alexander08}. This indicates the need for a subsequent phase of SMBH growth, which could be associated with a QSO phase. 

One puzzling issue about the proposed evolutionary connection between high-redshift QSOs and SMGs is only a small fraction of sub-millimeter detected galaxies (S$_{850} \gs 5$mJy), are identified as optically-luminous, M$_{\rm B } \leq -22$, $z\sim 1-3$ QSOs ($\sim 2$\%, \citealt{Chapman05,Wardlow11}), indicating that, if they are related, then the QSO and ULIRG phases do not overlap significantly.  Taken together, these various results provide support for an evolutionary link between high-redshift ULIRGs (SMGs) and QSOs, with a fast transition ($\sim 1$Myr) from the star-formation-dominated SMG-phase to the AGN-dominated QSO-phase (e.g.\ \citealt{Page12}), where rapid SMBH growth can then subsequently account for the present-day relation between spheroid and SMBH masses.  In this scenario, when a QSO {\it is} detected in the far-infrared/sub-millimetre it will be in the transition phase from an SMG to an unobscured QSO, making its physical properties (e.g.\ gas mass, dynamics, etc.) a powerful probe of the proposed evolutionary cycle (e.g.\ \citealt{Page04,Page11,Stevens05}). 

In this paper we present a study with the IRAM Plateau de Bure interferometer (PdBI)  of the cold molecular gas in far-infrared-selected ``transition'' candidate QSOs from the {\it Herschel}\,\footnote{Herschel is an ESA space observatory with science instruments provided by European-led Principal Investigator consortia and with important participation from NASA} H-ATLAS survey \citep{Eales10}. We detect $^{12}$CO\,(3--2) in both far-infrared bright QSOs studied, and compare the gas and kinematic properties of these to other $^{12}$CO-detected QSOs and SMGs to relate their evolution.  Throughout we adopt cosmological parameters from \citet{Spergel03} of: $\Omega_m =0.27$, $\Omega_\Lambda = 0.73$ and H$_0=71$\,km\,s$^{-1}$\,Mpc$^{-1}$.

%
%
\begin{figure}
\centerline{ \psfig{figure=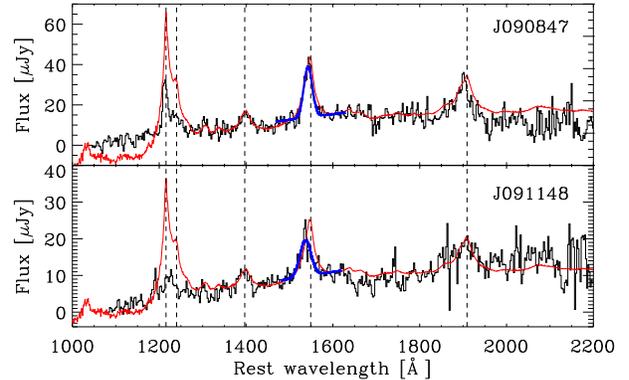,width=0.46\textwidth}}
\caption{The 2SLAQ UV rest frame spectra for both QSOs in our sample from \citet{Croom09}, with the 2QZ composite QSO spectrum overlaid (red; \citealt{Croom02}). From left-to-right, a dashed line indicates the rest wavelength of Ly$\alpha$, N{\sc v}, Si{\sc iv}, C{\sc iv} and C{\sc iii}] emission. Both QSOs, but especially J0911$+$0027, have relatively weak Ly$\alpha$\,1215 emission when compared to C{\sc iv}\,1549, suggesting the presence of significant quantities of neutral gas in their vicinity (see also \citealt{Omont96}). For each QSO we estimate the SMBH mass from the FWHM of the C{\sc iv}\,1549 line and the rest frame 1350\AA\ luminosity \citep{Vestergaard06}. The FWHM of the C{\sc iv} line is derived from the best fitting Gaussian and continuum model (shown in blue), which provides an adequate fit to the emission line in both cases: $\chi^2_{\rm{r}}=1.4$ for J0908$-$0034 and $\chi^2_{\rm{r}}=1.5$ for J0911$+$0027. The rest wavelength scale of the plots is based on the systemic redshifts derived from $^{12}$CO\,(3--2), see section \S~\ref{sec:analysis}.}
 \label{fig:2slaq}
\end{figure}

\section{Observations \& Data Reduction}
\label{sec:obsDR}

%
%
\begin{figure*}
\centering
\hspace{0.025\textwidth}
\subfloat{{\psfig{figure=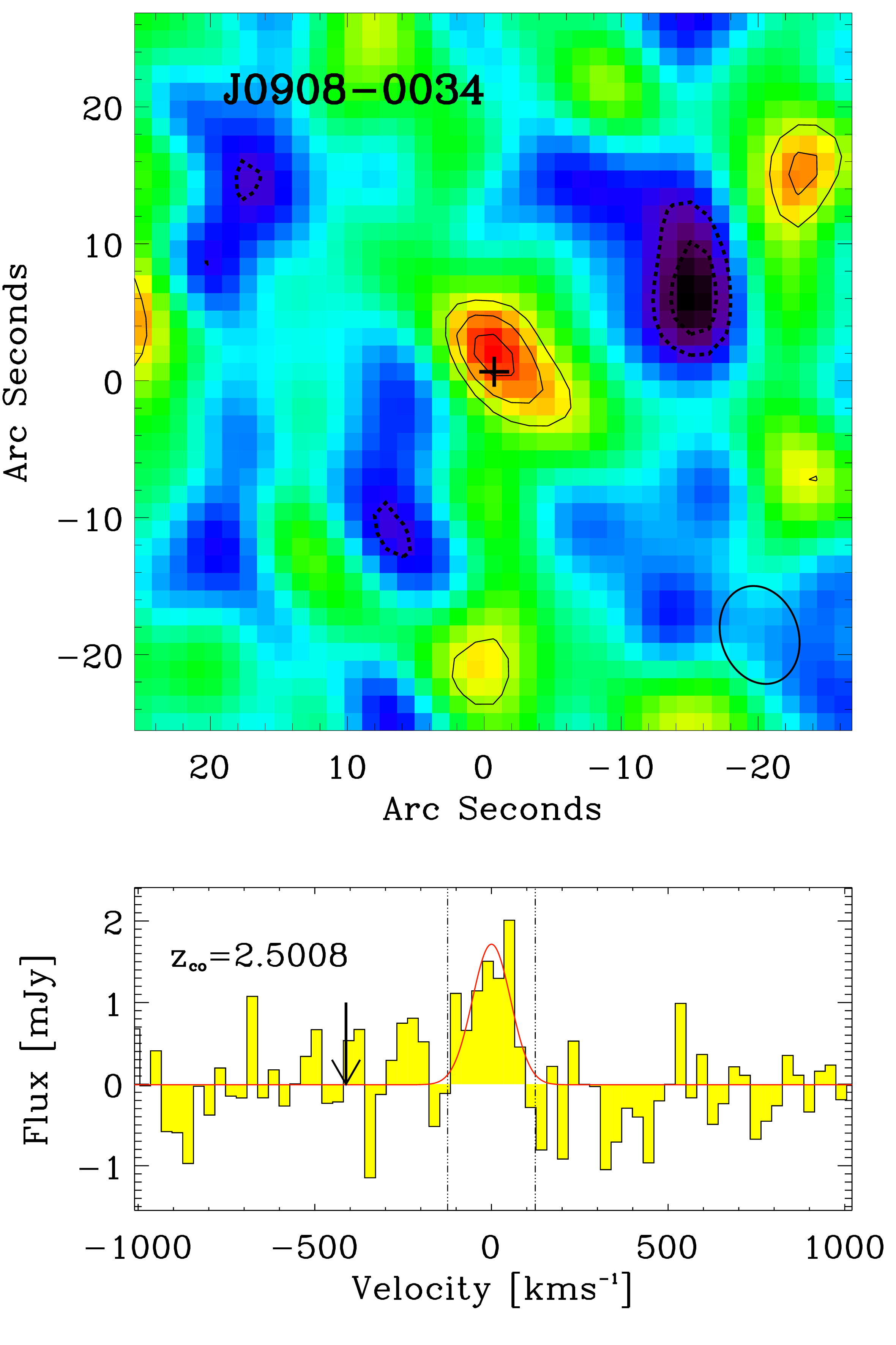,width=0.45\textwidth}}}
\hspace{0.025\textwidth}
\subfloat{{\psfig{figure=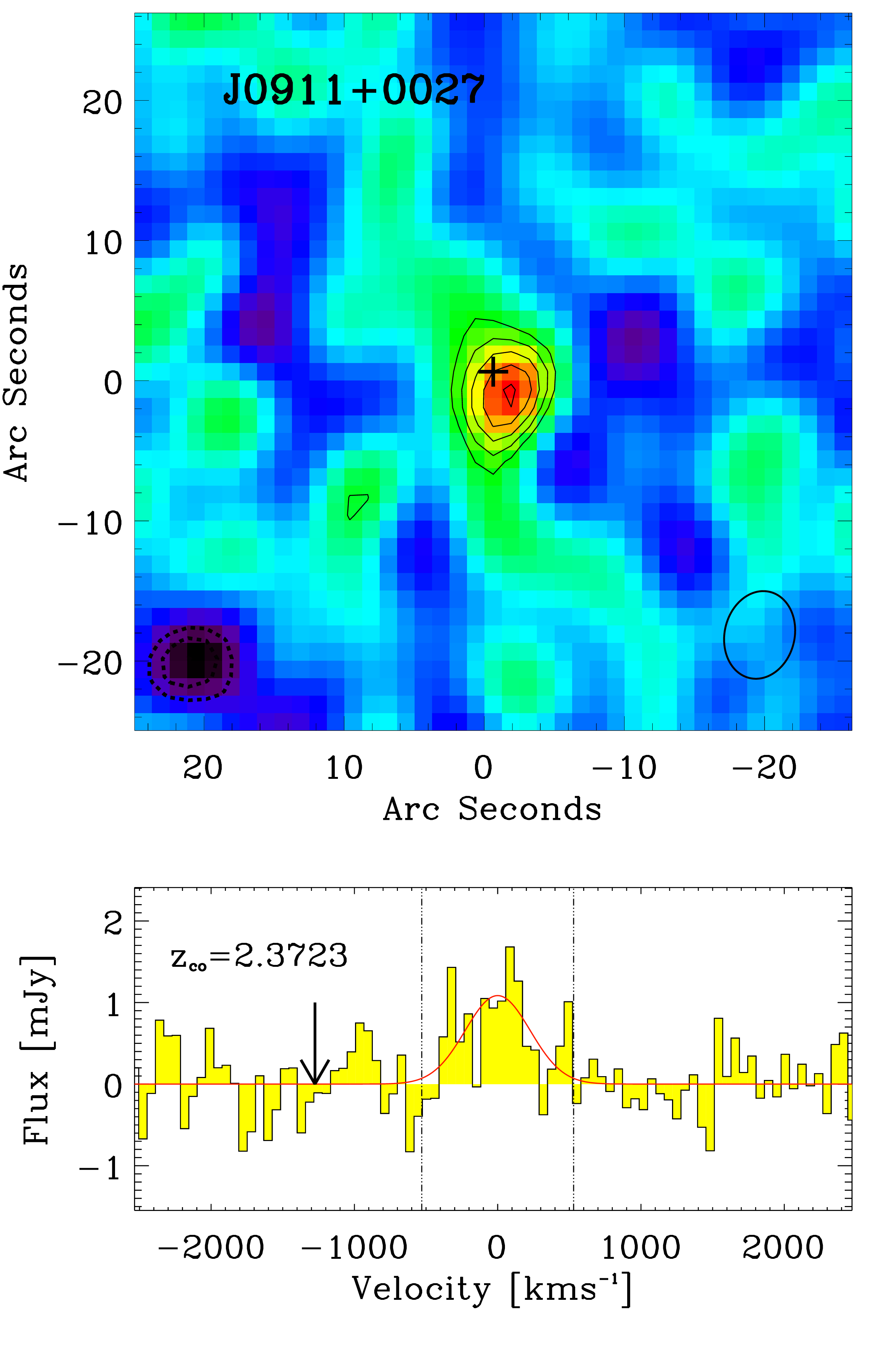,width=0.45\textwidth}}} 
\caption{ {\it{Upper}}: The velocity-integrated $^{12}$CO\,(3--2) emission for the two QSOs (summed over the regions marked in the spectra below). The contours represent signal-to-noise, and are spaced at $\pm 3,4,5,... \times \sigma$ (dashed contours are negative).   Significant $^{12}$CO\,(3--2) emission is detected from both QSOs and is well-centered on the optical positions of the QSOs, indicated by a cross (offset by 1--2$''$, which is within the primary beam).  Analysis of the channel maps shows both sources are unresolved with no evidence of velocity gradients, although the modest spatial resolution of our maps means that this is not a strong constraint.
{\it{Lower}}: Spectra showing the redshifted  $^{12}$CO\,(3--2) emission lines from both QSOs. The spectra were binned to a resolution of 10\,MHz and 20\,MHz respectively, and we overlay the best fitting  Gaussian plus continuum model  (the continuum emission in both QSOs is negligible). The redshift derived from $^{12}$CO\,(3--2) as well as the line flux and line widths derived from the Gaussian fits are reported in Table~2. We note that the FWHM of the J0908$-$0034 emission line is relatively narrow, at 125\,$\pm$\,25\,km\,s$^{-1}$, compared to 530\,$\pm$\,100\,km\,s$^{-1}$ for J0911+0027. In each spectrum the predicted postion of the $^{12}$CO\,(3--2) emission line, from the 2SLAQ UV spectrum, is indicated by an arrow. The two dot-dashed lines correspond to $\pm$FWHM, and show the region which was summed to produce the maps above.
}
 \label{fig:contour}
\end{figure*}

\subsection{Sample Selection}
\label{sec:sample}

Our aim is to study the molecular gas and dynamical properties of far-infrared bright QSOs to test their evolutionary link to SMGs, for which larger samples are now available \citep{Greve05,Bothwell12}.  We focus our study on potential transition (far-infrared bright) QSOs at the era when the activity in both the QSO and SMG populations peaked, $z\sim 2.5$.  Previous $^{12}$CO studies of unlensed far-infrared or sub-millimetre-bright QSOs~\citep{Coppin08b} have usually studied the most luminous QSOs, with SMBH masses 1--2 orders of magnitude larger than those seen in typical SMGs and correspondingly low space densities. Both of these factors make it hard to relate the properties of these extreme QSOs to typical SMGs (studies of lensed QSO samples, \citealt{riechers11c}, suffer from the difficulty of determining the true space densities of the sources being studied).  The limitations of these previous studies arose because of the lack of large samples of far-infrared/sub-millimetre detected QSOs; ~\citet{Coppin08b} selected a sample of QSOs which were later found to be bright in the sub-mm. But with the advent of wide-field far-infrared/sub-millimetre surveys with the {\it Herschel Space Observatory}~\citep{Pilbratt10} we can now select samples of QSOs with redshifts, space densities, far-infrared luminosities and most critically SMBH masses which are well-matched to those of the hypothesised descendents of typical SMGs.

%
%
\begin{table*}
{\small
\begin{center}
{\centerline{\sc Table 1: Observed Properties}}
\smallskip
\begin{tabular}{lcccccccccccc}
\hline
\noalign{\smallskip}
QSO                          & $\alpha$\,$_{\rm CO}$ & $\delta$\,$_{\rm CO}$    & S\,$_{250\mu{\rm m}}$ $^{a}$ & S\,$_{350\mu{\rm m}}$ $^{a}$ & S\,$_{500\mu {\rm m}}$ $^{a}$ & S\,$_{3{\rm mm}}$ \\
                                 & (J2000)                & (J2000)                                & (mJy)                    & (mJy)                    & (mJy)                     & (mJy)                  \\
\hline                                                                  
J0908$-$0034       & 09\,08\,47.18    & $-$00\,34\,16.6   &  9.0\,$\pm$\,6.6  & 26.2$\pm$\,8.1  & 15.0\,$\pm$\,9.0  & $<1.4$\\
J0911$+$0027       & 09\,11\,48.30    & $+$00\,27\,18.4  &  25.3\,$\pm$\,5.8  & 14.7\,$\pm$\,6.6  &  7.7\,$\pm$\,7.8  & $<1.0$ \\
\hline \hline
\end{tabular}
\vspace{-0.1cm}
\end{center}
\label{table:obs}

\begin{flushleft}
 \footnotesize{$^{a}$ J0908-0034 flux density values are after deblending sources. Fluxes extracted at the optical position of the QSO, before deblending, are 15.6\,$\pm$\,6.4\,mJy, 39.5\,$\pm$\,7.1\,mJy and 41.0\,$\pm$\,8.5\,mJy}
\end{flushleft}
}

\end{table*}

For our analysis we have therefore used the sample of far-infrared observed QSOs from \citet{Bonfield11} which are derived from the survey of the  9$^{h}$ H-ATLAS field \citep{Eales10}.  \citet{Bonfield11} extracted far-infrared fluxes from the SPIRE 250, 350 and 500-$\mu$m maps \citep{Griffin10,Pascale11} at the optical position of 372 QSOs from either SDSS \citep{Schneider10} or 2SLAQ \citep{Croom09} surveys across the $\sim 16$\,degree$^{2}$ field. A statistical background subtraction was performed on the measured fluxes for each QSO to account for confusion in the maps, see~\citet{Bonfield11}. Of the 372 QSOs, $\sim20\,\%$ are consistent with having L$_{\rm FIR}\geq 10^{12}$\,L$_\odot$, making them candidates for ``transition'' QSOs (e.g.\ between a far-infrared bright SMG and far-infrared faint unobscured QSO phase).  Restricting the redshift range to $z=2$--3 (to roughly match the redshift peak of SMGs, Chapman et al.\ 2005), gives 29 QSOs.  We then estimated SMBH masses for these QSOs from the extrapolated rest frame 5100-\AA\ luminosities (e.g.\ Wandel et al.\ 1999) to select those with SMBH masses of M$_{\rm BH}\leq 3\times 10^8$\,M$_\odot$, comparable to SMGs \citep{Alexander08}.  These selection criteria resulted in nine targets and from this list we selected six QSOs for a pilot project to determine their gas masses and dynamics through the detection of $^{12}$CO\,(3--2) emission with PdBI.  We estimated systemic redshifts for these QSOs using the measured wavelengths from Gaussian fits to the  Si{\sc iv}\,1397, C{\sc iv}\,1549 and  C{\sc  iii}]\,1909 emission lines in their spectra (Fig.~\ref{fig:2slaq}). We then adjust our estimates by the weighted mean velocity offsets of these lines from $^{12}$CO\,(3--2), for the $^{12}$CO-detected QSOs in \citet{Coppin08b} (we find Si{\sc iv}\,1397 provides the best estimate of the systemic redshifts).  We note that, as previously seen by \citet{Omont96}, roughly $\sim 80$\% of these far-infrared bright QSOs show very weak Ly$\alpha$ emission. There may be a relation between far-infrared luminosity and weak Ly$\alpha$, but its physical origin is still unknown. 

This paper presents the PdBI observations of the first two QSOs from this sample: J0908$-$0034 (RA: 09\,08\,47.18, Dec: $-$00\,34\,17.9, J2000; $z_{\rm UV}=2.5073$) and J0911$+$0027 (RA: 09\,11\,48.38, Dec: $+$00\,27\,19.7, J2000, $z_{\rm UV}=2.3697$).

\subsection {PdBI Observations}
\label{sec:pdbi}

We used the six-element IRAM PdBI in compact (D) configuration to search for redshifted $^{12}$CO\,(3--2) emission from the QSOs. We tuned the correlator {\sc WideX} to the frequency of redshifted $^{12}$CO\,(3--2)  from the estimated redshifts derived from their UV emission lines (Fig.~\ref{fig:2slaq}): 98.593GHz and 102.619GHz, for J0908$-$0034 and  J0911$+$0027 respectively. The advantage of {\sc WideX} is that its 3.6-GHz (dual polarisation) spectral coverage, at a fixed channel spacing of 1.95MHz, corresponds to a velocity range of $\sim$\,10,000\,km\,s$^{-1}$, sufficient to identify the $^{12}$CO\,(3--2) line even if the UV lines are significantly offset from the systemic redshift.  We obtained a total on-source observing time of 6\,hr per source between 2010 April 26 and 2010 May 4. The overall flux density scale was calibrated to 3C\,273, with observations of J0906+015 for phase and amplitude calibration.  Receiver bandpass calibration was performed on 0923+392.  The data were calibrated in the {\sc gildas} software package and a naturally weighted data cube produced for each QSO. In this configuration the synthesised beam for natural weighting is $6.3''\times 5.0''$ at a position angle (PA) of $166^{\circ}$ for J0908$-$0034, and $7.3''\times 5.6''$ at a PA of $19^{\circ}$ for J0911$+$0027.

\subsection{Far-infrared Luminosities and SMBH Masses}

For our analysis we require more accurate estimates of the far-infrared luminosities and SMBH masses of our QSOs. To estimate far-infrared luminosities of the QSOs in our sample, we 
exploit the {\it Herschel} SPIRE 250, 350 and 500-$\mu$m imaging.  For each 
QSO, we deblend multiple sources in the vicinity, using the 250$\mu$m map to identify nearby sources.  For 
J0911$+$0027 there is a bright 250$\mu$m source centered at the optical position of the QSO which is clearly isolated, but J0908$-$0034 has significant (\,40\%--60\% at 250--500$\mu$m) contamination from a galaxy $\sim$30$''$ away.  We report the 
final deblended photometry for each source in Table~1.  To derive the 
far-infrared luminosity for each QSO, we fit a modified black body 
spectra to the SPIRE photometry, adopting a dust emissivity, 
$\beta\,=\,1.6$ and fixing the temperature of the modified black body at 
T$_{\rm d}\,=\,40\,K$, (the average dust temperature of $z\sim$\,2--3 
QSOs, \citealt{Beelen06} ) and calculate the far-infrared luminosity 
(L$_{\rm FIR}$) by integrating the best-fit SED between (rest-frame) 
8--1000\,$\mu$m.  We note that if 
we instead allow the characteristic dust temperature to vary in the SED 
fit, we see at most a $\sim$45\% change in L$_{\rm FIR}$, consistent within our 
error estimates. The modified black body fits confirm the L$_{\rm{FIR}}$ for both QSOs is $\sim (3-4 \pm 1) \times10^{12}$\,L$_\odot$ (Table~2). The QSOs are thus significantly fainter in the FIR than the QSO sample of Coppin et al. (2008), L$_{\rm{FIR}} = (7.5\pm1.5) \times 10^{12}$\,L$_\odot$, and also at the fainter end of the SMGs in Bothwell et al. (2012), L$_{\rm{FIR}} = (4.8\pm 0.6) \times 10^{12}$\,L$_\odot$ .  The implied star-formation rates\footnote{SFR (M$_\odot$yr$^{-1}$) = $1.7\times10^{-10}$\,L$_{\rm{FIR}}$ (L$_\odot$) \citet{Kennicutt98}, following a Salpeter IMF over a mass range 0.1--100M$_\odot$ } (SFR) are 500--700\,M$_\odot$\,yr$^{-1}$.  

%
%
\begin{table*}
{\small
\begin{center}
{\centerline{\sc Table 2: Physical Properties}}
\smallskip
\begin{tabular}{lcccccccc}
\hline
\noalign{\smallskip}
Source                  & $z_{\rm CO}$     & S$_{\rm CO}\Delta\nu$   & FWHM$_{\rm CO}$  & L$^\prime_{\rm{CO\,(3-2)}}$  & L$_{\rm FIR}$ & M$_{\rm gas}$ & M$_{\rm BH}$   \\
                            &  & (Jy\,km\,s$^{-1}$)        & (km\,s$^{-1}$)     &  (10$^{10}$\,K\,km\,s$^{-1}$\,pc$^2$) & (10$^{12}$\,L$_\odot$) & (10$^{10}$\,M$_\odot$) & (10$^8$\,M$_\odot$)      \\            
\hline                                                                                                                                                                                                                                                       
J0908$-$0034      & 2.5008\,$\pm$\,0.0002 & 0.23\,$\pm$\,0.04 &  125\,$\pm$\,25 & 0.77\,$\pm$\,0.15 & 3.1\,$\pm$\,0.8 & 0.77\,$\pm$\,0.15 & 1.6$^{+0.3}_{-0.2}$   \\   
J0911$+$0027     & 2.3723\,$\pm$\,0.0005 & 0.62\,$\pm$\,0.10 & 530\,$\pm$\,100  & 1.87\,$\pm$\,0.33 & 3.5\,$\pm$\,0.8 & 1.9\,$\pm$\,0.3 & 2.5$^{+1.4}_{-0.9}$  \\

\hline\hline
\end{tabular}
\vspace{-0.1cm}
\end{center}
\label{table:phy}
}
\end{table*}

We calculated SMBH masses for both QSOs from the FWHM of their C{\sc iv}\,1549 emission lines and their rest frame 1350-\AA\ continuum luminosities (L$_{1350}$), following Eq.~7 from~\citet{Vestergaard06}. This relation is calibrated to the results of reverberation mapping, however the geometry of the broad line region (producing the emission lines) is poorly constrained and may bias virial mass estimates to low values~\citep{Jarvis06,Fine11}.  We derive the FWHM from a Gaussian fit to the C{\sc iv} emission line in the 2SLAQ spectra (Fig.~\ref{fig:2slaq}),  measuring  $4500\pm 340$\,km\,s$^{-1}$ and  $5900^{+1600}_{-1300}$\,km\,s$^{-1}$ for  J0908$-$0034 and J0911$+$0027 respectively, and determine L$_{1350}$ from the $g$-band SDSS magnitudes of $g=21.41\pm 0.05$ for  J0908$-$0034 and $g=21.64\pm 0.05$  for J0911$+$0027. As we do not see strong evidence of reddening in the 2SLAQ spectrum of either QSO we do not consider the effects of dust extinction on our estimates of L$_{1350}$ (Fig.~\ref{fig:2slaq}). From these we estimate SMBH masses of $\sim 2\times 10^{8}$\,M$_\odot$ for both QSOs\footnote{We estimate errors by bootstrap resampling the 2SLAQ spectrum, with replacement, and re-fitting a Gaussian model.} (Table~2). We caution that estimates from the C{\sc iv} emission line can over predict the SMBH mass by a factor of 2--5, when compared to estimates from H$\alpha$~\citep{Ho12}. In the absence of a more robust mass estimator, we simply acknowledge that the SMBH masses of our sources may be over estimated.

\section{Analysis \&  Results}
\label{sec:analysis}

We detect $^{12}$CO emission near the optical postion of each QSO (offset by 1--2$''$) and at frequencies close to those expected for redshifted $^{12}$CO\,(3--2). For each QSO, we fit a combination of a Gaussian and a uniform continuum to the spectrum corresponding to the peak signal-to-noise, $S/N$, pixel in each map and report the resulting redshift, line width and line flux in Table~2, as well as overplotting the fit on Fig.~\ref{fig:contour}.  We detect no significant continuum emission at 3mm in either QSO (3-$\sigma$ limits given in Table~1), which is consistent with no contribution from synchrotron emission to the far-infrared luminosities of these systems. The quoted 1-$\sigma$ errors were obtained from a bootstrap error analysis on the model fit to the data.  The resulting values of $\chi^2$ indicate good fits to both emission lines: $\chi^2_{\rm{r}} =1.0$ for J0908$-$0034 and $\chi^2_{\rm{r}} =1.1$ for J0911$+$0027.  Previous results have found $\sim$ 25\% of SMGs display a double-peaked emission line, with a velocity difference between peaks $\gs$\,500\,km\,s$^{-1}$ \citep{Bothwell12}, indicating either a merger or disk-like kinematics. To test for this we attempt to fit a double Gaussian model to the spectra in Fig.~\ref{fig:contour}. We find it produces a negligible improvement in $\chi^2$, and conclude that single Gaussian fits are sufficient to describe the line profiles of both QSOs.

We collapse the spectral cube over the frequency range corresponding to $\pm$\,FWHM of the emission line to create the maps of the  $^{12}$CO\,(3--2) emission shown  in Fig.~\ref{fig:contour}.   These maps are continuum subtracted, although as we noted above no significant continuum is detected in either source.  We now discuss the $^{12}$CO\,(3--2) emission line properties of each QSO in more detail.

\noindent{\it J0908$-$0034} ---
The integrated $^{12}$CO\,(3--2) emission is detected at a $S/N$ of 5.5$\sigma$. The $^{12}$CO emission line is relatively narrow and so we bin to a frequency resolution of 10\,MHz ($\sim$\,30\,km\,s$^{-1}$), as shown in Fig.~\ref{fig:contour}. We find the $^{12}$CO\,(3--2) emission line to be well fit by a single Gaussian with $\rm{FWHM} = 125\pm25$\,km\,s$^{-1}$, and a velocity-integrated flux of  S$_{\rm CO}\Delta\nu = 0.23\pm0.05$\,Jy\,km\,s$^{-1}$; see Table~2. From the peak of the Gaussian distribution, we determine a $^{12}$CO redshift of $z = 2.5008\pm0.0002$. We compared $z_{\rm CO}$ with the wavelength of emission lines in the 2SLAQ spectrum of this QSO (Fig.~\ref{fig:2slaq}). We found the UV lines to be blue-shifted by $\Delta v \sim$\,100--500\,km\,s$^{-1}$, relative to z$_{\rm CO}$, with Ly$\alpha$ and N{\sc v} emission lines having the largest offset. Systematic offsets of UV emission lines are not unusual, and are usually attributed to resonant scattering of the emission line, due to material in the line of sight to the AGN~\citep{Steidel10}. At the spatial resolution of our D-configuration PdBI observations we do not find any evidence for spatially resolved $^{12}$CO\,(3--2) emission, and channel maps show no sign of a velocity gradient across the source.

\noindent{\it J0911$+$0027} ---
The integrated $^{12}$CO emission is detected at a significance of 7.2$\sigma$. The $^{12}$CO\,(3--2) spectrum is binned to a frequency resolution of 20\,MHz  ($\sim$\,60\,km\,s$^{-1}$), as shown in Fig.~\ref{fig:contour}. We find the $^{12}$CO\,(3--2) emission line to be well fitted by a single Gaussian distribution with $\rm{FWHM} = 530\pm100$\,km\,s$^{-1}$, and a velocity integrated flux of S$_{\rm CO}\Delta\nu = 0.62\pm0.10$\,Jy\,km\,s$^{-1}$; see Table~2. We determine a $^{12}$CO redshift of $z = 2.3723\pm0.0005$ from the line fit.  The rest frame UV emission lines in the 2SLAQ spectrum of this QSO are blue-shifted by $\Delta v \sim$\,500--1500\,km\,s$^{-1}$, relative to $z_{\rm CO}$ (Fig.~\ref{fig:2slaq}). As with J0908$-$0034, Ly$\alpha$ and N{\sc v} emission lines display the largest offsets, but strong absorption in Ly$\alpha$ and N{\sc v} may affect the accuracy of the Gaussian fit to the emission lines. At the low spatial resolution of our D-configuration observations there is no evidence for spatially resolved $^{12}$CO\,(3--2) emission, and the channel maps do not show any indication of velocity gradients across the line.

We now compare our observations of these two far-infrared-luminous, but otherwise fairly typical,  QSOs with the more luminous (and more massive) far-infrared bright QSOs previously studied, as well as with the SMG population.  In particular, we investigate what the observed properties of our QSOs can tell us about their relationship to SMGs.

\subsection{Gas Mass}
\label{sec:gasmass}

%
%
\begin{figure}
 \centerline{  \psfig{figure=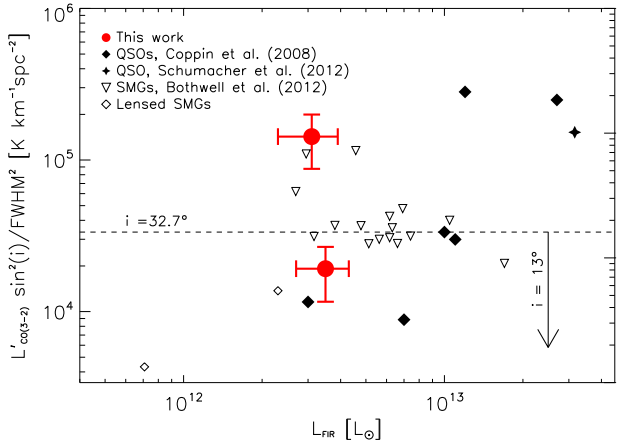,width=3.5in} }
\caption{The variation in L$'_{\rm CO(3-2)} \sin^2 i / {\rm FWHM}^2$, a proxy for molecular gas-mass fraction, as a function of far-infrared luminosity, for high-redshift QSOs and SMGs. We adopt a mean inclination for both populations, of 32.7$\degree$, appropriate for randomly orientated discs. For each sample we determine the median, but given the similarity of the results, we show only the QSO value (dashed line). Our proxy for gas mass fraction assumes the gas reservoirs in QSOs and SMGs have the same geometry, physical radius and mean inclination on the sky. We do not expect SMGs to suffer an inclination bias, but this may not be true for the optically-identified QSOs we study here, as the identification is based on their broad line properties. We indicate with an arrow the effect of adopting a mean inclination of 13$\degree$ for the QSO population, which may be more realistic, see \S\ref{sec:gasfraction}. Our proxy also assumes the population have the same line brightness ratio, see \S\ref{sec:gasmass}. A value of $r_{31} = 0.52$ is more appropriate for SMGs, and would further increase the separation between the implied gas mass fractions of SMGs and QSOs, with the later having lower gas fractions, and hence appearing more evolved. }
 \label{fig:fwhm2}
\end{figure}

%
%
\begin{figure}
 \centerline{
  \psfig{figure=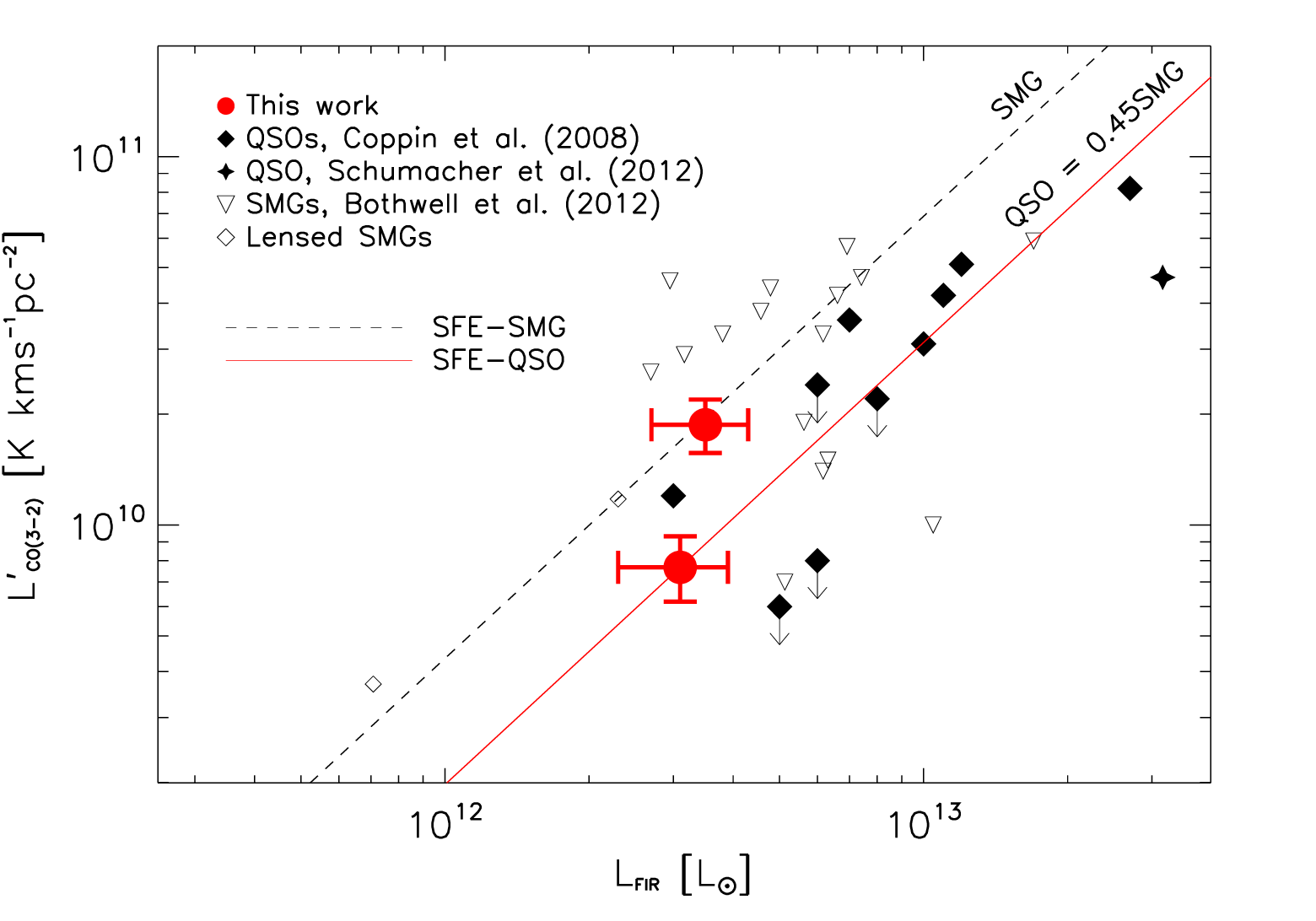,width=3.5in}}
\caption{$^{12}$CO\,(3--2) line luminosity (L$^\prime_{\rm CO\,(3-2)}$) versus far-infrared luminosity (L$_{\rm FIR}$) for far-infrared bright QSOs and SMGs.  We plot the best-fit line to the SMG population and we fix the gradient of the fit to the QSOs to that measured for the SMGs, ${\rm dL}^\prime_{\rm CO\,(3-2)}/{\rm dL}_{\rm FIR} = 1.2$ \citep{Bothwell12}. At a fixed L$_{\rm FIR}$ the QSOs have L$^\prime_{\rm CO\,(3-2)}$ which is a factor of $\sim$\,50\,$\pm$\,23\% lower than the SMGs, at a significance of 1.8-$\sigma$.  If real, this difference is in the sense that the far-infrared bright QSOs appear to have shorter gas consumption timescales, consistent with representing a subsequent evolutionary phase. If we take into account the difference in the line brightness temperature ratios, $r_{31}$, for the two populations the statistical significance of this difference increases to $3.1\sigma$.} 
 \label{fig:lcolfir}
\end{figure}

The  $^{12}$CO\,(3--2) line emission from each QSO provides information about the molecular gas mass of the system~\citep{solomon97}. We determine the $^{12}$CO\,(3--2) line luminosity, L$^\prime_{\rm CO\,(3-2)}$, following Eq.~3 from~\citet{SolomonVandenBout05}, deriving L$^\prime_{\rm CO\,(3-2)}=(0.77\pm 0.15) \times 10^{10}$\,K\,km\,s$^{-1}$\,pc$^2$ and L$^\prime_{\rm CO\,(3-2)}=(1.87\pm 0.33)\times 10^{10}$\,K\,km\,s$^{-1}$\,pc$^2$ for J0908$-$0034 and J0911$+$0027 respectively (Table~2).  

To estimate the total masses of the gas reservoirs in these QSOs we use the $^{12}$CO\,(3--2) luminosity to estimate the expected $^{12}$CO\,(1--0) luminosity and from that estimate the total gas mass.  For the first step we use the results from \citet{Riechers11d} and assume a typical line brightness temperature ratio, for the gas in QSOs, of $r_{31}$ = $L^\prime_{\rm CO\,(3-2)}$\,/\,L$^\prime_{\rm CO\,(1-0)} \sim 1$  (i.e.\  thermalised gas) to convert the $^{12}$CO\,(3--2) line luminosities of our QSOs, and the \citet{Coppin08b} sample, to  $^{12}$CO\,(1--0) luminosities. We caution that we are extrapolating the results of \citet{Riechers11d} to our sample of QSOs, and that in AGN dominated systems super-thermal ratios, i.e. $r > 1$, are not uncommon (e.g.\ \citealt{Papadopoulos08,Ivison12}). We then adopt a CO-to-H$_2$ conversion factor of $\alpha = 1$\,M$_\odot$( K\,km\,s$^{-1}$\,pc$^{2}$ )$^{-1}$ following \citet{Bothwell12} to convert L$^\prime_{\rm CO}$\,(1--0) to give the gas masses (Table~2). Note that $\alpha$ is denoted as a conversion to H$_2$, but the resulting H$_2$ mass is defined as the total H$_2$+He gas mass. In this manner we derive gas masses of M$_{\rm gas}=(0.77\pm 0.15) \times 10^{10}$\,M$_\odot$ and M$_{\rm gas}=(1.9\pm 0.3)\times 10^{10}$\,M$_\odot$ for J0908$-$0034 and J0911$+$0027 respectively. 

The gas masses of our far-infrared luminous QSOs are lower than the median gas mass of the extremely far-infrared luminous QSOs detected by~\citet{Coppin08b},  M$_{\rm gas}=(4.3 \pm 1.0)\times \,10^{10}$\,M$_\odot$, and they are also lower than the median gas mass of SMGs from~\citet{Bothwell12}, M$_{\rm gas}=(5.3 \pm 1.0)\times 10^{10}$\,M$_\odot$. We note that there is significant scatter around these averages, and the larger of our QSO gas masses is consistent within 3-$\sigma$. To allow a fair comparison, we have transformed the literature estimates of M$_{\rm gas}$ to use $\alpha = 1$. We also note that the bulk of the observations come from $^{12}$CO\,(3--2), although \citet{Bothwell12} use $r_{31}=0.52\pm 0.09$ in their conversion of L$^\prime_{\rm CO}$\,(3--2) to L$^\prime_{\rm CO}$\,(1--0), which is appropriate for SMGs due to the presence of an extended reservoir of cold gas. As we will discuss in  \S\ref{sec:discussion}, the lower gas masses we derive for our far-infrared QSOs are consistent with them having recently evolved from a previous-SMG phase.

\subsection{Gas Fraction}\label{sec:gasfraction}

One crude measure of the evolutionary state of a system is the fractional contribution of molecular gas to its total mass.  Hence, in the absence of significant replenishment of gas from external sources, we would expect the gas-mass fraction in SMGs to decline as they evolve, and thus in the evolutionary model we are testing, our QSOs ought to have lower gas-mass fractions than SMGs.

In Fig.~\ref{fig:fwhm2} we use a combination of observables which trace gas mass, L$'_{\rm CO(3-2)}$, and dynamical mass, ${\rm FWHM}^2/\sin^2 i$, as a proxy for gas-mass fraction, L$'_{\rm CO(3-2)} \sin^2 i / {\rm FWHM}^2$. In choosing this proxy we make two implicit assumptions: {\it (i)} that the gas reservoir in both SMGs and QSOs is distributed in the same manner {\it (ii)} the physical extent of the reservoir is similar in the two populations. Fig.~\ref{fig:fwhm2} shows that the far-infrared bright QSOs have similar values of L$'_{\rm CO(3-2)} \sin^2 i / {\rm FWHM}^2$ on average to SMGs.  To determine if this means that they have similar gas-mass fractions (potentially indicating they are not more evolved) we need to consider the two simplifying assumptions above.

As their sub-millimetre emission is optically thin it is believed that SMGs do not suffer from an inclination selection bias and so, if we are modelling their gas distributions as disc-like systems \citep{Swinbank11}, we can adopt a mean inclination of 32.7$\degree$, appropriate for randomly orientated discs~\citep{Bothwell12}. Recent studies comparing $^{12}$CO linewidths of QSOs and SMGs found no distinction between the populations, indicting that they have the same average inclination~\citep{Coppin08b}. As such, in our analysis we adopt the same inclination for QSOs and SMGs. However, optically-identified QSOs such as those studied here may have preferentially biased orientations, as to have identified them as QSOs we must be able to observe their broad-line regions. Even considering situations where the obscuring torus of the AGN is not orientated with the host galaxy, we still expect a bias in the inclination of the gas reservoir. This arises from consideration of when the line of sight to the broad line region is directly through the disk of the host galaxy. If we adopt a mean inclination of 13$\degree$, which has been previously suggested~\citep{Carilli06}, the QSOs sample has an $\sim 85\%$ lower gas mass fraction, on average, than the SMGs. The difference is significant at only $2.0\sigma$, but is in the sense expected if QSOs are more evolved systems.  

To convert L$'_{\rm CO(3-2)}$ to a total gas mass we must make assumptions about the temperature distribution within the gas. $^{12}$CO\,(3--2)  has an excitation temperature of 33\,K and a critical density of $5\times 10^4$\,cm$^{-3}$, meaning it traces warm and/or dense environments.  In QSOs it is thought that the gas is close to thermal equilibrium, meaning that the $^{12}$CO\,(3--2) transition traces the majority of the gas in the system \citep{Riechers11d}. \citet{Bothwell12} show that the $^{12}$CO(J\,,\,J$-$1) J\,$\geq$\,2 emission lines in SMGs are sub-thermally excited  (see also~\citealt{Harris10}), which is interpreted as an indication for the presence of multiple temperature components: a warm component associated with the star-forming regions and a cooler component associated with extended, quiescent, gas (see also \citealt{Danielson11}). Our proxy for the gas mass fraction assumes the conversion from L$'_{\rm CO(3-2)}$ to gas mass is the same for both the QSO and SMG populations, which is a conservative assumption. If instead we were to include the differences in line brightness temperature ratio ($r_{31}$) then the gas mass fraction is roughly doubled in SMGs, the offset between the populations in Fig.~\ref{fig:fwhm2} becomes larger and the statistical formal significance of the difference increases to $3.6\sigma$ (including a bias in the QSO inclination).  We conclude that although the estimated gas mass fractions of far-infrared bright QSOs appear similar to typical SMGs, this is subject to a number of assumptions about the kinematics, excitation, sizes and orientations of the two populations.

%
%
\begin{figure}
\centerline{
 \psfig{figure=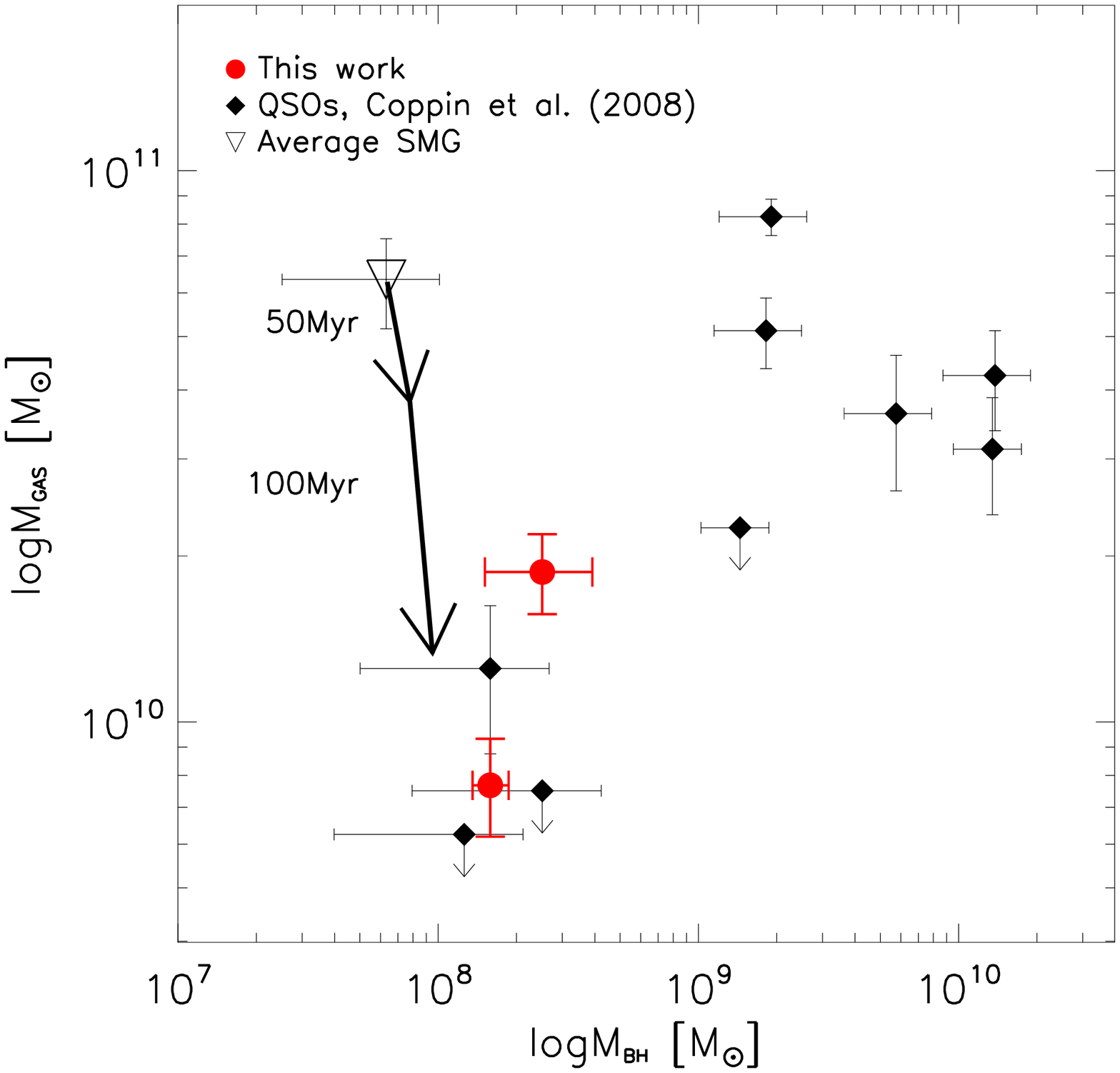,width=0.46\textwidth}}
\caption{A comparison of gas and SMBH masses for far-infrared bright QSOs and SMGs which have been observed in $^{12}$CO\,(3--2). The QSO points are taken from~\citet{Coppin08b} and the average SMG gas mass from \citet{Bothwell12}, and SMBH mass from \citet{Alexander08}. We show the predicted evolution of the gas mass and SMBH mass in a model SMG  in which star formation and SMBH growth occur in tandem for a period of 50 or 100\,Myrs.  
For the star formation we assume a constant rate of 500\,M$_\odot$yr$^{-1}$ \citep{Bothwell12}. The growth of the SMBH is calculated following Eq. 10 from~\citet{Alexander12}, and we assume it is accreting with an Eddington ratio of $\eta=0.2$ \citep{Alexander08}.  We conclude that the properties of the far-infrared bright QSOs in our sample, and the QSOs with M$_{\rm BH} < 10^9$\,M$_\odot$ from~\citep{Coppin08b}, are consistent with their recent evolution from an earlier SMG-like phase. The far-infrared bright QSOs with M$_{\rm BH} > 10^9$\,M$_\odot$ are attributed as statistical outliers in this evolutionary scenario, as their SMBH masses are amongst the largest seen in the local universe. }
 \label{fig:bhvgas}
\end{figure}

\subsection{Gas Consumption Time Scales}

A second important measure of the evolutionary state of SMGs and QSOs comes from their gas-consumption timescales:  M$_{\rm gas}/\rm SFR$ (or the inverse of this: the star-formation efficiency). In our evolutionary scenario the gas reservoir in SMGs is depleted by star formation as they transform from dusty systems to unobscured QSOs. Our `transition' FIR-bright QSOs are expected to represent a unique evolutionary phase, with prodigious star formation but lower gas masses, and thus shorter gas consumption timescales.

The ratio of the observables, $^{12}$CO\,(3--2) line luminosity, L$^\prime_{\rm CO(3-2)}$,  and far-infrared luminosity, L$_{\rm FIR}$, can be used as a proxy for the gas consumption timescale of a galaxy.  This ratio has the advantage that we do not hide any differences between populations in the assumptions about the conversion of $^{12}$CO\,(3--2) measurement to $^{12}$CO\,(1--0). We have therefore gathered samples of SMGs and FIR-bright QSOs which have been detected in $^{12}$CO\,(3--2), allowing a direct comparison of observable quantities. We  use literature values for the $^{12}$CO\,(3--2) luminosities of  QSOs \citep{Coppin08b,Schumacher12}, SMGs \citep{Bothwell12} and lensed SMGs \citep{Kneib05,Iono09,Swinbank10Nature,Danielson11}.  We plot our QSOs along with these comparison samples on Fig.~\ref{fig:lcolfir}.  We see that of our observations follow a similar trend to the literature QSO values. Combining our results with the QSOs sample, we find they display $\sim$\,50\,$\pm$\,23\% lower ratios of L$^\prime_{\rm CO(3-2)}/$\,L$_{\rm FIR}$ than SMGs, potentially indicating shorter gas consumption timescales. The statistical significance of this difference is only 1.8-$\sigma$, but we stress that this interpretation implicitly assumes the same value for  $r_{31}$ in the two populations.  If we use the proposed $r_{31}$ values for QSOs and SMGs (see \S\ref{sec:gasmass}) this would increase the difference between the two populations as the SMG gas masses will roughly double relative to the QSOs (significant at the 3.1-$\sigma$ level), owing to the presence of a reservoir of colder gas in the SMGs, which is not traced by the $^{12}$CO\,(3--2) transition and which is not seen in the QSOs \citep{Riechers11d}. Thus it appears that the far-infrared bright QSOs not only lack the extended, cool reservoir of gas seen in SMGs (e.g.\ \citealt{Ivison10L1L2,Ivison11EVLA}) but also have less warm/dense gas than SMGs, suggesting that they are more evolved.

\section{Discussion}
\label{sec:discussion}
In order to study the link between QSOs and SMGs we have selected two far-infrared-detected QSOs at $z\sim 2.5$ (the epoch of peak activity in SMG and QSO populations) whose SMBH masses ($\sim 2\times 10^8$\,M$_\odot$) are comparable to the average SMBH mass for the SMG population~\citep{Alexander08}. In contrast the majority of previously $^{12}$CO --detected (unlensed) far-infrared bright QSOs have M$_{\rm BH}\sim 10^9$--$10^{10}$\,M$_\odot$ \citep{Coppin08b}, so much larger than the average SMG M$_{\rm BH}$ that they cannot have recently evolved from a typical SMG (as pointed out by ~\citealt{Coppin08b}). It is precisely this mismatch in SMBH masses which our study addresses. Given their FIR luminosities and SMBH masses our new sample is a more accurate representation of the typical far-infrared bright QSO population, and hence can be used to test an evolutionary link between SMGs and QSOs, through their gas and dynamical masses, as traced by their $^{12}$CO\,(3--2) emission.

We show in Fig.~\ref{fig:bhvgas} the distribution of estimated gas and SMBH masses for our two QSOs, as well as the average SMG from \citet{Bothwell12} and the QSO sample (including limits) from \citet{Coppin08b}. To investigate how galaxies might evolve with time we follow \citet{Coppin08b} and overlay a simple evolutionary model based on gas consumption and SMBH growth timescales. We take the average SFR of SMGs from \citet{Bothwell12} as 500\,M$_\odot$\,yr$^{-1}$ and use this to calculate the reduction in their gas mass with time. In this simple model we assume a constant SFR and do not include mass loss due to winds. By selection our QSOs have SMBH masses similar to the average SMGs, but to estimate the growth of the central SMBH on the gas consumption timescale we follow Eq.~10 from ~\citet{Alexander12}. In line with the measured properties of SMGs, we limit the growth to an Eddington ratio, $\eta=0.2$~\footnote{ $\eta=0.2$ indicates fast black hole growth. Within uncertainties it is feasible the growth is Eddington limited, $\eta=1$.}, and an efficiency of 0.1 \citep{Alexander08}, this then predicts the estimated growth in the SMBH mass in parallel to the depletion of the gas reservoir. 

In Fig.~\ref{fig:bhvgas} we plot tracks showing the evolution in the expected gas and SMBH masses for the descendents of SMGs after 50 and 100\,Myrs.  As can be seen, after 100\,Myrs the SMGs in this model are expected to have almost completely depleted their massive, cold gas reservoirs and will have gas masses comparable to those we detect in the far-infrared luminous QSOs.  At the same time their SMBHs will have grown by $\sim 50$\%, resulting in masses similar to those seen in our target QSOs. As can be seen from Fig.~\ref{fig:bhvgas}, the rate of growth of the SMBH and the depletion of the gas reservoir can link the properties of a typical SMG to those seen in our far-infrared bright QSOs around $\sim$\,100\,Myrs later. 

Are all SMGs likely to go through a QSO-phase? If these three $z\sim$\,2--3 populations (i.e.\ S$_{\rm 850\mu m} \gs $\,5\,mJy SMGs and QSOs with SMBH with masses of $\gs $\,10$^8$\,M$_\odot$, which are both far-infrared-bright and -faint) are uniquely related through a simple evolutionary cycle, then the product of their respective space densities and lifetimes should be similar. On average we expect the SMG lifetimes to be comparable to their gas depletion timescales (Fig.~\ref{fig:bhvgas}) or $\sim$\,10$^{8}$\,yrs \citep{Swinbank06b,Hickox12} and their volume density at $z\sim 2-3$ is $\sim 10^{-5}$\,Mpc$^{-3}$ (e.g.\ \citealt{Wardlow11}).  In comparison, the QSO volume density at $z\sim 2-3$ (with SMBH masses above $\sim 10^8$\,M$_\odot$) is $\sim 10^{-6}$\,Mpc$^{-3}$ \citep{Croom09} and estimates of their lifetimes are $\gs 10$\,Myrs \citep{Martini01,Hosokawa02}. As noted by a number of previous authors, the space densities and duty cycles of these populations are thus consistent with all SMGs subsequently transforming into a QSO. If they transform through a far-infrared bright QSO phase how long can this be? From~\citet{Chapman05} and ~\citet{Wardlow11} we estimate that far-infrared bright QSOs have a volume density at $z\sim 2-3$ of $\sim 10^{-7}$\,Mpc$^{-3}$. By comparison to the SMG and QSO lifetimes this implies a duration for this transition phase of just $\sim 1$\,Myr. Estimates of lifetimes and space densities for each population are uncertain at $> 2\times$, and so a qualitative agreement is reasonable. If the QSO influences its host through winds and outflows with characteristic velocities of $\sim 1000$\,km\,s$^{-1}$ (e.g.\ \citealt{Harrison12}) then the estimated duration of the far-infrared bright phase would allow these winds to reach kpc-scales, comparable to the likely extent of the gas reservoirs and star-formation activity within these systems (e.g. \citealt{Tacconi08,Ivison11EVLA}). This may help explain the short duration of the `transition' phase, although it may be exhaustion of the gas reservoir by star formation which is allowing these winds to propagate freely. 

\section{Conclusions}

The main conclusions from our study are:
\begin{itemize}
\item We have used IRAM PdBI to search for redshifted $^{12}$CO\,(3--2) emission from two far-infrared bright QSOs at $z\sim 2.5$ selected from the H-ATLAS survey. These QSOs were selected to have SMBH masses of M$_{\rm BH}\leq 3\times 10^8$\,M$_\odot$, which are more comparable to typical SMGs than those in samples of high-redshift far-infrared luminous QSOs previously detected in $^{12}$CO.  Our observations detect $^{12}$CO\,(3--2)) emission from both QSOs and we derive line luminosities of L$'_{\rm CO(3-2)} = (0.77\pm 0.15) \times 10^{10}$\,K\,km\,s$^{-1}$\,pc$^2$ and L$'_{\rm CO(3-2)} = (1.87\pm 0.33)\times10^{10}$\,K\,km\,s$^{-1}$\,pc$^2$ for J0908$-$0034 and J0911$+$0027 respectively (Table~2).

\item Comparing our FIR-bright QSOs (and similar systems from the literature) with SMGs, we find that the QSOs have similar values of L$'_{\rm CO(3-2)} \sin^2 i / {\rm FWHM}^2$, our proxy for gas mass fraction, to SMGs. However this is subject to a number of assumptions. If we consider that QSOs have a biased inclination angle of 13$\degree$, as has been suggested~\citep{Carilli06}, then the QSOs have an $\sim$85\% lower gas mass fraction, at a significance of 2.0--$\sigma$. In the absence of gas replenishment this is consistent with QSOs being more evolved systems. Furthermore, adopting the appropriate line brightness ratios for each population roughly doubles the gas mass in SMGs, increasing this difference further, and to a significance of 3.6--$\sigma$. So while we see no evidence for evolution in the gas fraction, we require spatially-resolved $^{12}$CO\,(1--0) observations to test this conclusively.

\item By comparing the gas consumption timescales, estimated from L$'_{\rm CO(3-2)}$\,/\,L$_{\rm FIR}$, for far-infrared bright QSOs with SMGs, we find that the QSOs have $\sim$\,50\,$\pm$\,23\% lower consumption timescales, at a significance of 1.8--$\sigma$.  Adopting appropriate line brightness ratios, of r$_{31}$=1 for QSO, and r$_{31}$=0.52 for SMGs strengthens this conclusion to 3.1--$\sigma$.  We conclude that far-infrared bright QSOs have a lower mass of warm/dense gas (probed directly through $^{12}$CO\,(3--2)). Combined with previous results~\citep{Riechers11d}, showing that QSOs also lack an extended, cool reservoir of gas seen in SMGs, we interpret this as evidence that the far-infrared bright QSOs are at adifferent evolutionary stage than typical SMGs. In our evolutionary scenario this is consistent with far-infrared bright QSOs being in `transition' from an SMG to a QSO.

\item We show that the gas and the SMBH masses in far-infrared bright QSOs and SMGs are consistent with a model where SMGs transform into QSOs on a timescale of $\sim 100$\,Myrs.  Furthermore, the relative volume densities and expected durations of the SMG and QSO phases are consistent with all SMGs passing through a subsequent QSO phase, and we estimate that the likely duration of the far-infrared bright QSO phase is just $\sim 1$\,Myr.  We note that if necessary this duration is still sufficient to allow the QSO to influence the star formation and gas reservoirs across the full extent of the host galaxy through 1000-km\,s$^{-1}$ winds and outflows.

\end{itemize}

The scale of this study is too small to single handedly prove or disprove an evolutionary link between SMGs and QSOs. However the data we have obtained provides further evidence supporting the idea originally proposed by \citet{Sanders88} that these populations are linked by an evolutionary sequence. We have compared $^{12}$CO\,(3--2) detected QSOs and SMGs through a number of observable quantities, and find the timescales for gas depletion, and SMBH growth, needed to link SMGs to these sources are consistent. 

\section*{Acknowledgments}
This work is based on observations carried out with the IRAM PdBI. IRAM is supported by INSU/CNRS (France), MPG (Germany) and IGN (Spain). We thank Chiara Feruglio, Stephen Fine and Tom Shanks for help.

JMS acknowledges the support of an STFC studentship. IRS, AMS and DMA acknowledge financial support from the STFC. IRS acknowledges the support from a Leverhulme Senior Fellowship. 

KEKC acknowledges support from the endowment of the Lorne Trottier Chair in Astrophysics and Cosmology at McGill, the Natural Science and Engineering Research Council of Canada (NSERC), and a L'Or\'{e}al Canada for Women in Science Research Excellence Fellowship, with the support of the Canadian Commission for UNESCO. 

The Herschel-ATLAS is a project with Herschel, which is an ESA space observatory with science instruments provided by European-led Principal Investigator consortia and with important
participation from NASA. The H-ATLAS website is http://www.h-atlas.org/

\bibliography{ref.bib}

\end{document}